\documentclass[pss,a4paper]{sa}
\usepackage{times}
\usepackage{sa}
\usepackage[latin1]{inputenc}
\usepackage[T1]{fontenc}
\usepackage{amssymb,amsmath}
\usepackage{color}
\usepackage[]{graphicx}
\begin{document}
\renewcommand{\copyrightyear}{2012}
\DOIsuffix{theDOIsuffix}
\Volume{4} \Issue{1} \Copyrightissue{01} \Month{01} \Year{2006} 
\pagespan{1}{}
$$\Receiveddate{\sf zzz} \Reviseddate{\sf zzz} \Accepteddate{\sf zzz}
$$\Dateposted{\sf zzz} 



\definecolor{lightgray}{gray}{0.85}

\hspace{368px}\colorbox{lightgray}{\color{black} \large{Physics}}\\

\title[thesis title]{Study of the Influence of High Electric Field
Variations on\\ Cosmic Ray Flux detected by the
ARGO-YBJ Experiment}

\author[Irene Bolognino]{Irene Bolognino\inst{1,2}}

\address[\inst{1}]{Dipartimento di Fisica, Università di Pavia, via Bassi 6, 27100 Pavia, Italy}
\address[\inst{2}]{Istituto Nazionale di Fisica Nucleare, sezione di Pavia, via Bassi 6, 27100 Pavia, Italy \\irene.bolognino@pv.infn.it}
\begin{abstract}
This paper is an overview of the author's PhD thesis results \cite{tesi}.  
ARGO-YBJ is an extensive air shower detector located at Yangbajing (Tibet, China) at 4300 m a.s.l. It is made by a full coverage carpet plus a guard ring (total surface $\sim6700\;m^{2}$) of Resistive Plate Chambers grouped into 153 units called {\it{clusters}}.
The experiment has two different operation modes. The former, the {\it{scaler mode}}, counts the number of events with particle multiplicity $\geq1$,$\geq2$,$\geq3$,$\geq4$ allowing to reach the lowest energy threshold of the detector (few GeVs). The latter, the {\it{shower mode}}, measures coordinates and arrival time of each particle hitting the carpet for a complete shower reconstruction at an energy threshold of few
hundreds of GeVs.
Due to the high sensitivity of scaler mode it becomes very important remove all environmental parameter effects from scaler counts. 
The careful study of cosmic ray variations with environmental va\-ria\-bles (such as atmospheric pressure, temperature, local radioactivity) allowed the correct estimation of the significance of signals registered in scaler mode.
Episodes of counting rate variations both in scaler and shower mode in correspondence to high electric field variations are discussed, together with the study of temporal and spatial characteristics of cosmic ray showers during thunderstorms.
\end{abstract}

\maketitle                   




\renewcommand{\leftmark}
{I. Bolognino: Study of the Influence of High E. Field
Variations on CR Flux detected by the
ARGO-YBJ Exp.}

\section{Introduction}

The study of cosmic ray flux variations due to strong atmospheric discharges is basic not only to understand cosmic ray physics but also to know more about atmospheric phenomena. ARGO-YBJ is a full coverage extensive air shower array located at high altitude, 4300 m. a.s.l., and designed for very
high energy gamma-ray astronomy and cosmic ray observations with energies ranging
from a few GeV up to the PeV region. The detector works in two modes: the scaler and
the shower mode (for details see next section). In scaler mode, an unexpected performance is
observed concerning the lowest multiplicity channel which seems to be affect by natural local radioactivity in addition to atmospheric pressure and RPC gas temperature \cite{aielli2008}. The first part of this work is in fact dedicated to the study of the influence of 222-Radon daughter in air on scaler 1 (C1) counting rates. Moreover about half of C1 counts seems to be due to soil radioactivity which has been estimated through analytical ways.
Finally the best correction function of the data on scaler mode has been worked out and applied to the data.\\
The experimental site is provided with two electric field mills (Boltek EFM-100) and a Vaisala Weather Transmitter in order to study the effects of high electric field variations on the shower rate and development. The second part of this work is therefore dedicated to the study, both in scaler and in shower mode, of temporal and spatial characteristics of cosmic ray showers during thunderstorm. In particular this study is focused on the search for possible common characteristics for the several lightning investigated events.

\section{The ARGO-YBJ Experiment}

ARGO-YBJ is an extensive air shower detector located at the International Laboratory of Yangbajing in Tibet (P.R. China) at 
4300 m a.s.l. It has been optimized to detect cosmic rays and 
$\gamma$ radiation through a full coverage carpet plus a guard ring (total surface $\sim6700\;m^{2}$) of Resistive Plate Chambers (RPCs) \cite{webargo}. 
The basic module is called ``cluster'' (5.7 x 7.6 m$^{2}$) divided 
into 12 RPCs, which defines the base unit for the data acquisition system \cite{RPCaielli2006}. 
ARGO-YBJ has been designed to operate in two independent acquisition modes:
the scaler mode and the shower mode. In the former technique, which allows to reach the lowest energy threshold 
(E $\sim$ 1 GeV), the total counting rates of each cluster are recorded every 0.5 s, without information on arrival direction and spatial distribution of the detected particles. 
Four low multiplicity channels in each cluster are implemented for event multiplicities from
$\geq 1$ to $\geq 4$ and the counts from different pads of the same 
cluster are put in coincidence in a narrow time window (150 ns).
The mean measured counting rates for each cluster are respectively $\sim$ 40 kHz, 
$\sim$ 2 kHz, $\sim$ 300 Hz and $\sim$ 120 Hz.\\
Scaler moder is useful to study flaring phenomena such ad gamma-ray bursts,
solar flares, forbush decreases and also to monitor the 
influence of meteorological effect, mainly pressure and gas temperature.\\
The shower mode is based on the requirements that a minimum number of pads
(20) must be fired in the central carpet with the proper space-time pattern (time window 420 ns). The
time resolution ($\sim$ 1 ns), the small temporal (the pad, each RPC is made by 10 pads) and spatial (the strip, each pad is made by 8 strips) pixels, the high granularity and the full coverage allow imaging of the shower front with unprecedented details. The position and time of any fired pad will be recorded to reconstruct
the shower parameters (core position, arrival direction and shower size). 
The shower data are used to study cosmic ray physics at energy threshold of about 1 TeV, 
very high energy gamma astronomy at $E_{th}\sim$ 300 GeVs and gamma-ray bursts. \\

\section{Study of the Influence of Natural Radioactivity on the Detector Counting Rates}

Due to their high sensitivity, scaler counts are influenced by environmental 
parameters, mainly represented by ambient pressure, temperature and natural radioactivity.
Remove all these effects becomes very important in order to get a signal as ``clean'' as possible and characterize events with
high accuracy. \\
The dominant meteorological effect is induced by atmospheric pressure which can causes changes in total counting rate amplitudes up to a few per cent.
Scalers from $\geq 2$ (called also C2) to $\geq 4$ (C4) are well corrected through a fit performed with a
two-dimensional function, as a first approximation linear in P (pressure) and T (gas temperature)
given by the following expression \cite{aielli2008}:

\begin{equation}
  C(P,T)\simeq C_0 [1-\mu(P-P_0)][1+\beta(T-T_0)]
  \label{aiellicorrection}
\end{equation}

where P$_0$ and T$_0$ are the pressure and temperature reference values
for counting rate C$_0$, and $\mu$ and $\beta$ are respectively the barometric
and the thermal coefficients. $\mu$ $=$ 0.9-1.2 \%/mbar and $\beta$ $=$ 0.2-0.4 \%/$^\circ$C were obtained \cite{cappa}, depending on the cluster considered and
the specific experimental conditions. These results agree with those of the other experiments.
Eq. \ref{aiellicorrection} does not fit well on C1 (scaler 1) instead, because it gives a lower barometric coefficient  $\mu$ $=$ 0.3-0.5 \%/mbar.
Looking for this unexpected behaviour of
C1 the hypothesis that radon daughters affect that counts has been taken into account. The other scalers are not influenced because the probability that two or more $\gamma$-rays, produced by radon daughters, hit the same cluster within 150 ns is negligible.

\subsection{Radon-222 gas in Air}

Yangabajing soil contains $^{238}$U, $^{232}$Th radioactive nuclides which
decay in the ground originating radioactive families, and $^{40}$K isotope. In particular radium-226 ($^{226}$Ra), belonging 
to the uranium chain, gives $^{222}$Rn which enters the buildings being an inert gas
and thus extremely volatile. It has an half time of 3.82 days and produces $^{214}$Pb and $^{214}$Bi
that in turn emit $\gamma$-rays able to influence C1 counting rate \cite{daughters}. Even if $^{222}$Rn is originated in the soil, once it flows inside the experimental hall it can be considered as an independent phenomenon respect to soil radioactivity. Through the {\it{secular equilibrium}} concept is possible to asses radon daughters concentration simply measuring the parent one. The secular equilibrium is a situation in which the activity, i. e. the number of decays that occur in a second, of the parent is equal to the daughter one.
It can only occur in a radioactive decay chain if the half-life of the daughter is much shorter than the half-life of the parent. In this case the half-lives of $^{214}$Pb and $^{214}$Bi are, respectively, 27 and 20 minutes that are much shorter than 3.82 days of $^{222}$Rn. The secular equilibrium is reached after $\sim$ 2-3 hours.
$^{222}$Rn concentration (C$_{Rn}$) is monitored every 30 mins by two Lucas cells located in the centre and in the North side of the RPCs carpet.
A Lucas cell is a type of scintillator which only $^{222}$Rn enters because of three filters. Once it is inside radon starts decaying originating its daughters \cite{nero}. 
The two monitors gave, and are still giving, two different values of C$_{Rn}$. At the 
centre C$_{Rn}$ has an average value of $\sim$ 500 Bq/m$^3$ while at the North side is $\sim$ 4 times higher but in some periods overlaps the centre one. Radon gas concentration at the North side of the experiment hall is therefore extremely variable.
These two different concentrations suggest that radon enters the ARGO-YBJ hall from soil and cracks, 
mainly on the North side and exits through doors and windows with an ease dependent on ventilation and atmospheric conditions \cite{icrc}. Two additional campaigns using passive nuclear track detectors confirmed Lucas cell results. 

\subsubsection{Monte Carlo Simulations}

Monte Carlo simulations were performed by using FLUKA code \cite{fluka1, fluka2} simulating a cluster of 43 m$^2$ with a volume of air of width 23 m, length 17 m and variable deep from 3 mm to 4 m which is the height of ARGO-YBJ roof. Six millions of gammas were launched in all directions covering all the energy spectrum of 
the isotopes in air. 500 Bq/m$^3$, 1000 Bq/m$^3$ and 3000 Bq/m$^3$ were considered as average values of radon gas concentration, supposed uniformly distributed in the hall air, and values of the equilibrium factor of 0.5, 0.7 and 1. An equilibrium factor \cite{equibfact} equal to 1 means that no atoms are attached or deposited on surfaces, but all the daughters are in air, contrary to 0 which shows that all the daughters are attached or deposited on surfaces.
Taking into account all the possible heights at which the gamma can be produced and the possibility that C$_{Rn}$ and equilibrium factor vary during the time, the result of Monte Carlo simulations shows an influence of natural radioactivity of $\sim$ 1-3\%  on scaler 1 counting rate. Thus a Bq/m$^3$ of $^{222}$Rn gas concentration in the hall air gives roughly 1 Hz on the scaler 1. The error of  simulations is about 3\%.

\subsection{Data Analysis}

The data analysis to assess radon gas contribution to scaler 1 (C1) counts is based on the
following time series x(t): radon gas in air C$_{Rn}$ (measured at the North side
and at the detector centre), atmospheric pressure P, detector gas temperature T, and scaler
counts C1, C2, C3 and C4 \cite{nostropaper}. These time series has been normalized (scaled) in order to make them comparable through the formula: $x(t)_{scld} = \frac{x(t)-<x(t)>}{\sigma_{x(t)}}$,
where x(t) is the experimental data at time t, <x(t)> and $\sigma_{x(t)}$ are the arithmetic mean and the
standard deviation, respectively, both calculated all over the examined period (typically of a
few weeks). The normalized series x(t)$_{scld}$ maintain the same behaviour as the original
one x(t) and have mean 0 and standard deviation 1.
Data series were chosen belonging to different seasons of 2010 and several clusters
were analysed depending on their position inside the experimental hall: North side, in the centre and at the South side.
Since radon data are collected every 30 minutes, the data of scaler counts, temperature
and pressure were averaged over the same period.
The analysis was performed with two different methods: the
{\it{method of linearization}} and the {\it{proportional method}}.

\subsubsection{Method of linearization}

The cosmic ray contribution to
C1 counts is assumed, as a first approximation, to depend linearly on both atmospheric pressure P and detector gas
temperature T, according to equation:

\begin{equation}
  C1(t)= a + bP(t) + cT (t) + C1_{RESIDUE}(t)
  \label{linearization}
\end{equation}

where {\it{a}} represents the mean contribution from both cosmic rays and the detector background
(supposed mostly influenced by the soil natural radioactivity) assumed constant as a first
approximation, while the residual term C1$_{RESIDUE}$ represents how much radon can influence C1 counting rate and thus it is expected to have a linear dependence to radon gas 
concentration. The correlation coefficient between C$_{Rn}$ and C1$_{RESIDUE}$ is likely to
be high when the C1 time variations produced by environmental phenomena other than radon
are negligible.
The results of this method show that clusters located in the centre of the carpet have a better correlation coefficient (on average $\sim$ 0.8) than those at the North (0.5-0.6) and at the South side (0.4-0.6) of the detector.
In particular the linear regression coefficients between C1$_{RESIDUE}$ and C$_{Rn}$ are calculated. 
They represent how much radon is influencing C1 counting rate. Values
range around 1 Hz/(Bq/m$^3$), depending on the cluster position and on the analysed period.
The percentage of the influence of natural radioactivity on scaler 1 counts is obtained  
through the division between the counting rate of the C1$_{RESIDUE}$ and the C1 one.
The values obtained for the analysed periods range from 1\% to 3\%. This result
agrees with Monte Carlo simulation one.

\subsubsection{Proportional Method}

According to the proportional method, C1 counts are considered as the sum of three signals (Eq. \ref{propeq}):
the cosmic ray contribution, namely $\gamma_1$, the radon contribution kC$_{Rn}$ and the detector background, Bck, mostly influenced by the soil natural radioactivity and assumed constant as a first approximation. 
\begin{equation}
  C1(t)=\gamma_1(t)+kC_{Rn}(t)+Bck
  \label{propeq}
\end{equation}
where k quantifies how much radon gas in air is increasing the C1 counts. 
Considering the dependence on pressure (P) and gas temperature (T),
cosmic rays are contributing to all the multiplicity channels with the same proportion, it is possible to
calculate $\gamma_1$ value using the higher multiplicity channels: $\gamma_1(t) = h_2 \gamma_2(t) = h_3 \gamma_3(t) = h_4 \gamma_4(t)$.
Since $^{222}$Rn is negligible on C2, C3 and C4, as a first approximation, the average proportion between the
scaler rates due to cosmic rays can be quantified as $h_n = \langle {\frac{[C1(t)-Bck]}{Cn(t)}} \rangle$ 
with n=2, 3 and 4, and $\gamma_2$ $=$ C2, $\gamma_3$ $=$ C3 and $\gamma_2$ $=$ C4 counts, in absence of other physical phenomena influencing the higher multiplicity channels.
The radon contribution to the C1 is obtained by Eq \ref{propeq}: $C1_{NET}= kC_{Rn}(t) = C1(t) -\gamma_1(t)-Bck$.
The maximization of the correlation coefficient between C$_{Rn}$ and C1$_{NET}$ allows the assessment of the
Bck value. Moreover, the proportion between the two series, C1$_{NET}$ and C$_{Rn}$, shows how much
radon gas in air is affecting the C1 channel.
It is important to notice that, because of the properties of the normalized series and assuming Bck as a constant, C1$_{NET}$ variations can be also calculated by subtracting from C1, C2 or C3 or C4 equivalently:\\
  $C1_{NET}(t)_{scld} = C1(t)_{scld} - C2(t)_{scld} = C1(t)_{scld} - C3(t)_{scld} = C1(t)_{scld} - C4(t)_{scld}$\;,
when all the series are in the normalized. 
This equation allows to assess the radon influence on C1 even when
C$_{Rn}$ was not still measured in the ARGO-YBJ hall (i.e. before 2009).
The results of this method are consistent with the linearization method ones.
It is important to notice that this method is
evidencing the Bck term of the order of 20 $\pm$ 5 kHz (depending on periods and clusters),
in agreement with previous results \cite{aielli2008}. 

\subsubsection{The Best Data Correction Formula}

Taking into account all the effects that influence scaler 1 the right correction formula is: 
\begin{equation}
  C1(t)= \gamma_{1}(t) + \mu \: \gamma_{1}(t) (P(t)-P_0) + \beta \: \gamma_{1}(t) (T(t)-T_0)+ kC_{Rn}(t) + Bck
  \label{mycorrection}
\end{equation}
$\gamma_1$(t), $\mu$, $\beta$, k and Bck are obtained through $\chi^2$ minimization performed with MINUIT ROOT cernlib.  \\
The analysis was performed for different time periods and for all the clusters of the ARGO-YBJ carpet obtaining value of
$\gamma_1$ of about (20 $\pm$ 1) kHz, $\mu$ parameter ranges from -1.0 \%/mbar to -0.85 \%/mbar with an error of $\pm$ 0.3 \%/mbar, depending on the analysed cluster and time period, while $\beta$ value is unchanged.
The new $\mu$ value agree with C2, C3 and C4 ones.
k parameter results of about 0.8-1.3 with an error of 0.2, using as $C_{Rn}$(t) the one
at the centre of the carpet: this confirms that $C_{Rn}$(t) at the centre can represent the average value over all
the RPCs carpet in addition to confirm the radon influence on C1. 
Finally the value of the soil radioactivity
ranges from 18 kHz to 25 kHz with 1 kHz as error.
The soil radioactivity value obtained confirms the hypothesis that half of C1 counts are not due to cosmic ray contribution \cite{{aielli2008}}: a dedicated study is currently in
progress.

\section{Theory of Lightnings and Thunderstorm Phenomena}
The study of cosmic ray physics is very important to understand atmospheric phenomena due to thunderstorms and
lightnings. The theory of {\textbf{runaway breakdown}} \cite{gurevich} (adopted in 1992) points out that cosmic rays significantly affect the state of
thunderclouds and apparently have a strong effect on both
ordinary lightning discharges and the new types of giant
atmospheric discharges between clouds and the ionosphere.
The phenomenon of runaway breakdown (RB) is based on
specific features of the interaction between fast particles
and matter. The braking force F acting on an energetic particle as it traverses matter is determined by the ionization
losses. In the nonrelativistic regime, the braking force is proportional to the molecular density and inversely 
proportional to the electron energy $\epsilon$. For $\epsilon \geq$ 1.5 MeV, the braking force reaches
a minimum F$_{min}$ and then slowly increases logarithmically.
The strong decrease in frictional scattering gives rise
to the possibility of accelerated electrons in a thundercloud's electric fields. Indeed, in a constant electric field E
that exceeds the critical field E$_c$, given by E$_c =$ F$_{min}$/e, e is the electron charge, an
electron with a sufficiently high energy $\epsilon > \epsilon_c \simeq$ mc$^2$E$_c$/(2E)
is continuously accelerated by the electric field. These electrons are called runaway electrons and may in turn 
generate more free electrons with $\epsilon > \epsilon_c$.
As a result, an exponentially growing
runaway avalanche can occur. \\
The condition E $>$ E$_c$ alone is however insufficient for RB.
The presence of fast ``seed'' electrons, having energies
above the critical runaway energy of 0.1-1 MeV, is also
necessary. Even more important, the spatial scale of the
electric field must substantially exceed the characteristic
length l$_a$ needed for the exponential growth of a runaway
avalanche. In the atmosphere of a thunderstorm, the characteristic
sizes of clouds are always much greater than l$_a$ and fast seed electrons are also plentiful, effectively
generated by cosmic rays. If the flux of fast seed electrons are nonuniform in space, the discharge spreads in the plane
orthogonal to the direction of the electric field and always remains inside a cone with an angle of $\theta_c = 2 \sqrt{\frac{D}{l_a}}$. 

\section{Study of the Influence of Atmospheric Electric Field on Cosmic
Ray Flux with the ARGO-YBJ Experiment}

ARGO-YBJ is equipped with two electric field mills Boltek EFM-100, located on the roof at the North and South side.
The response time of the two EFM-100 is 0.1 seconds while the electric field range was changed from $\pm$ 20 kV to $\pm$ 100 kV during the shift turn on November
2011, which resulted more appropriate on the basis of the electric field variations measured during the first data taking period. 
During good weather conditions electric field on the ground usually varies between $\pm$ 2.5 kV/m while during thunderstorms it grows almost at $\pm$ 100 kV/m. 
Due to the high variations reached by the atmospheric electric field, saturations of EFM-100 devices can be observed during thunderstorms.\\
Moreover the Vaisala Weather Transmitter WXT520 was installed on the roof in 2010. A lightning detector Boltek LD-250 is going to be installed on the roof after a testing period inside the experiment hall. Data analysis were performed both in scaler and operation mode considering data from 2008 to 2012 with particular interest for data from 2010, year in which both electric field monitors and Vaisala Weather Transmitter were in operation.

\subsection{Data Analysis in Scaler Operation Mode}

More than 150 events were observed during the last four years. These events show a similar pattern with enhancements in C1 and C2 and, in some cases, decreases in C3 and C4. The trend of the electric field depends on the kind of the storm: in some cases only a positive device saturation was registered, while in others only a negative one was present and some events have both values of saturation. Figure \ref{24maggio} shows three events registered on 24th May 2010. Three thunderstorms occurred in the Yangbajing sky at about 6:00 a.m., 9:00 a.m. and 1:00 p.m. and lasted respectively 50, 50 and 85 minutes. 
Electric field has the same behaviour in all the three atmospheric disturbances and changes in dramatic way. 
In fact at the beginning it shows a small increase respectively of 2 kV/m, 6 kV/m and 6 kV/m. These signals mean that several thunderclouds were approaching the sky above ARGO-YBJ and some lightnings were occurring inside them. Since the electric field hasn't high variations lightnings were intra-cloud and/or quite far. Then electric field drops until the value of device saturation -20 kV/m: this saturation means that thunderstorm was above the detector. The following dramatic change of sign, from -20 kV/m to +20 kV/m shows that a (or more) strong lightning stroke occurred, most probably a cloud to ground lightning leader \cite{hugh} followed by some return strokes. Return strokes, which occur some milliseconds after a lightning leader, could be responsible of the achievement of the positive charge of the atmospheric electric field \cite{multiple}.\\
The three strong enhancements observed in C1 and C2 counting rates mean that more particles reached the detector because of the runaway breakdown which originates electron avalanches.
Electrons having low energies are therefore be able to reach the ground without be absorbed by the atmosphere and can be thus detected by ARGO-YBJ, increasing the counting rate.
The decreases of C3 and C4 suggest that particles hitting the detector during a storm are more spreaded. As seen in the previous section electron avalanche can develop
in a cone and not in a straight line and besides lightning strokes occur mainly crosswise.  These considerations lead to the conclusion that particles fall down more spreaded during thunderstorms.
The significances of the signals shown in Fig. \ref{24maggio} were calculated on scalers once that all the effect of environmental parameters were removed. The study about natural radioactivity, described above, is therefore very important in order to have a signal as ``clean'' as possible and thus to determine the most accurate value of the significance of the observed signal. Eq. \ref{mycorrection} was used to perform the right correction of the analysed data.  \\
\begin{vchfigure}[!h]
\includegraphics[scale=0.6]{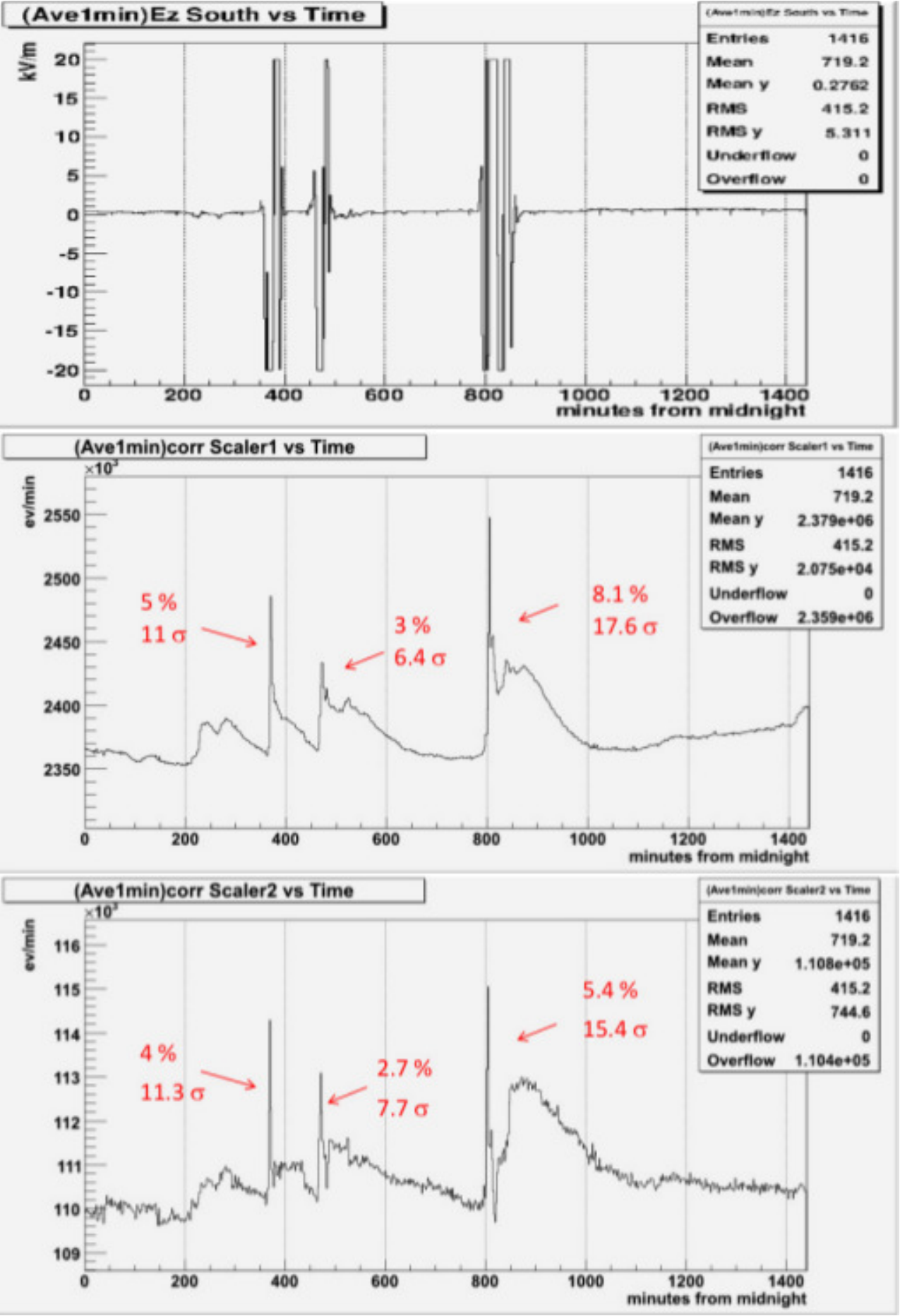}
\includegraphics[scale=0.6]{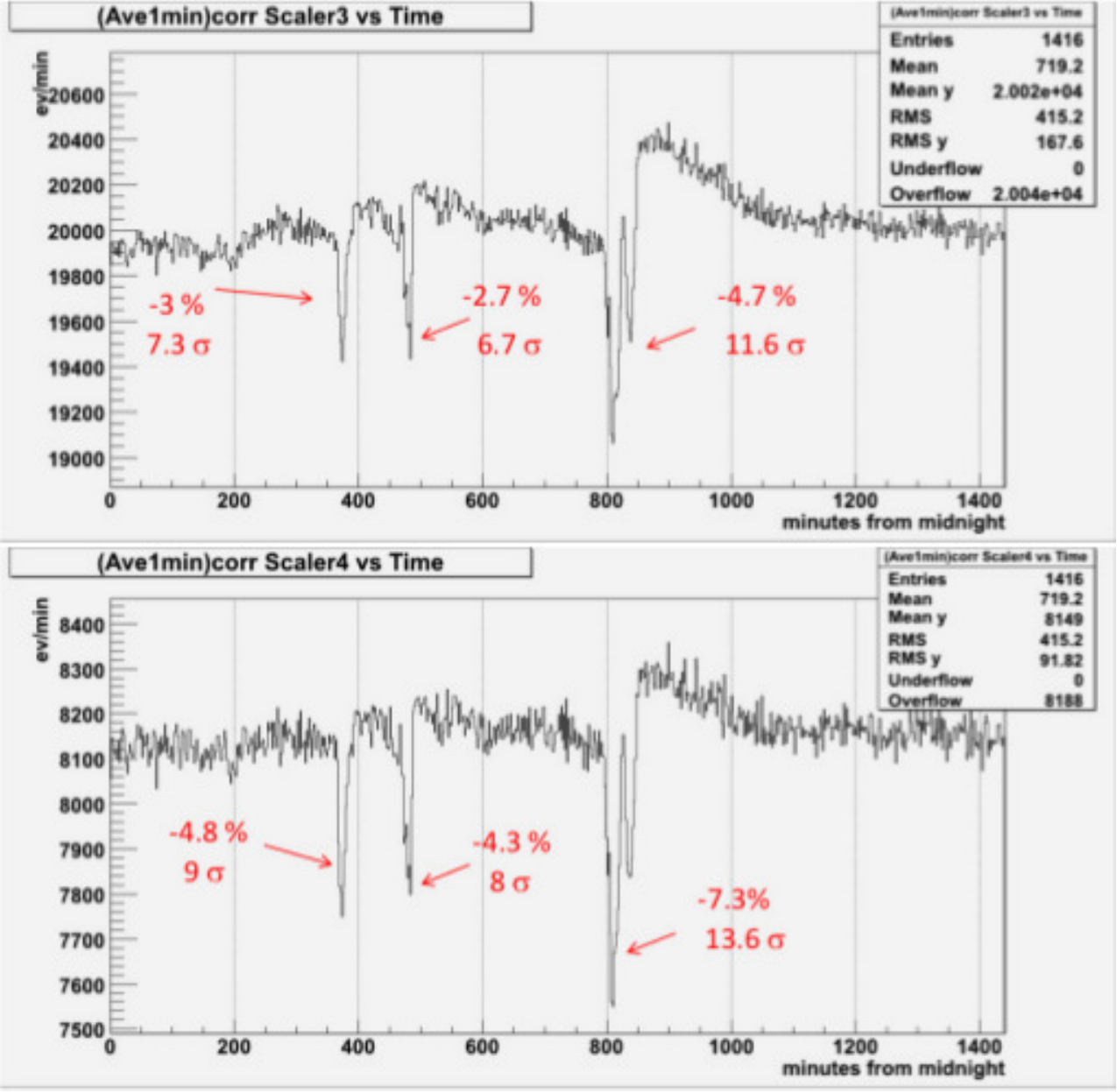}
\caption{Atmospheric electric field (top plot), C1 and C2 (left side) and C3 and C4 (right side) registered on 24th May 2010. In correspondence of
saturations of EFM-100, high enhancements were observed in C1 and C2, whilst C3 an C4 showed decreases. Significances of signals are shown on the plots. The x-axis unit is time in minutes from midnight while y-axis unit is V/m (electric field plot) or events in a minute (scaler plots).}
\label{24maggio}
\end{vchfigure}
In some events any effect are observed on C3 and C4. This is probably due to the high electric field which reaches value of $\pm$ 100 kV/m and induced a huge number of avalanche electrons that balance the effect of a lower particle density.\\
Even if events have a similar patter, because of thunderstorms are different from each other, any classification of the detected events is not possible. In fact even if two or more electric fields have the same behaviour, i.e. only positive, only negative, both positive and negative, runaway breakdown can occur at different times during these thunderstorms and moreover the type and the intensity of lightnings is not surely the same. Furthermore atmospheric conditions can be very different from one event to another. 
Two peaks belonging to two thunderstorms with similar electric field can therefore not to be compared to each other. This feature is also confirmed by the absence of published papers about a classification of cosmic ray flux variations in correspondence of atmospheric electric field ones.\\
Finally it is important to say that all scaler voltages were checked during these phenomena in order to leave out any possible electronic malfunction.

\subsection{Data Analysis in Shower Operation Mode}

Strong enhancements are registered also in shower mode in correspondence to high variations of the atmospheric electric field. Peaks in the shower mode are generally more significant than scaler mode ones. It is important to point out that
only in some cases there are both signals in scaler and in shower mode. Some of the events present in scaler mode could not be registered by the shower mode because of the lack of data
due to the dead time between two consecutive runs or, more frequently, because of the trigger condition is not reached (at least 20 pads are fired on the central carpet within a time window of 420 ns). 
The analysis was performed taking into account 5 cuts of primary energy, i.e. selecting the number of hits with energy E $\sim$ 1-2 TeV, E $\sim$ 2-4 TeV, E $\sim$ 4-6 TeV, E $\sim$ 6-8 TeV and E $>$ 8 TeV \cite{moon}.  
The comparison of the conical fit $\chi^2$ (through data are reconstructed) under the peak to the background one suggests that particles under peak are more scattered and thus their showers have a cone angle larger than background ones.
This result was confirmed also by the {\it{compactness}} parameter analysis. Compactness is a conical fit parameter which shows the average of the distance 
between the pads got on fire by the core. The barycentre of the shower is weighed over the pads. \\
In conclusion, when the atmospheric electric field has great variations shower particles are more scattered due to runaway breakdown mechanisms born inside thunderclouds. 
On the other side no changes in the particle arrival times were registered.\\
Knowing that the distribution of the arrival particles changes spatially during a thunderstorm it is worth asking if also the arrival direction changes. For
this purpose the azimuth ($\phi$) and the zenith($\theta$) angles were analysed. All the analysed events show an increase of showers coming from zenith angles from 0\textdegree\: to 20\textdegree\:, whereas the azimuth angle changes according to the feature of the analysed thunderstorm.

\section{Conclusions}

The goal of this work was to study the influence of high electric field variations on cosmic ray flux detected by the ARGO-YBJ experiment.
In order to perform this kind of analysis it has been necessary to take into account all the effect of the environmental parameters on the counting rates. The study concerned
the lowest multiplicity scaler, which is thus the most sensible, and which is also affected by the influence of the local natural radioactivity besides
the influence of ambient pressure and RPCs gas tem\-pe\-ra\-tu\-re. It was found that 1\%-3\% of counts on C1 are due to the radioactivity in air. Higher influence seems to be caused by the gamma rays emitted by radionuclides in the soil but a further measurements campaign is needed in order to confirm the result. Taking into account the natural radioactivity effects, the scaler 1 correction formula has been worked out.
Once that all scalers were corrected the analysis of the influence of high electric field variations on cosmic ray flux was performed both in scaler and shower operation mode.\\The effect of lightnings is to accelerate and spread the incoming particles. Depending on the balance between these two effects, enhancements in the rate of C1, C2 and in shower mode and/or decreases in the higher multiplicity rates (C3 and C4) can be observed.\\
Even if thunderstorm events are different to each other a common feature concerning the shower zenith angle was found. An increase of showers ar\-ri\-ving within 
20\textdegree\: has been observed. This can be due to the higher frequency of vertical or quasi-vertical electric discharges in the atmosphere. 
Observed differences in the azimuth component could be due to the different incoming direction of thunderstorms or to specific features of the events which are different from each other.\\
The study didn't highlight any difference in shower particles arrival times.

\bibliographystyle{ieeetr}

\end{document}